\documentclass[prd, aps, superscriptaddress, preprintnumbers, floatfix, nofootinbib]{revtex4}
\usepackage[dvips]{graphicx}
\usepackage{epsf}
\usepackage{amsmath}
\usepackage{amssymb}
\usepackage{varioref}
\usepackage{float}
\usepackage{soul}

\def\be{\begin{equation}}
\def\ee{\end{equation}}
\def\bea{\begin{eqnarray}}
\def\eea{\end{eqnarray}}

\usepackage{color}
\usepackage{ulem}

\usepackage[unicode=true,pdfusetitle,
 bookmarks=true,bookmarksnumbered=false,bookmarksopen=false,
 breaklinks=false,pdfborder={0 0 1},backref=false,colorlinks=true]
 {hyperref}
\hypersetup{
 linkcolor=blue, citecolor=magenta, urlcolor=red, filecolor=blue}

\def\eq#1 { \begin{equation} #1 \end{equation} }
\def\eqn#1{ \begin{eqnarray} #1 \end{eqnarray} }

\def\gui{\textcolor{red}}
\def\rod{\textcolor{green}}

\begin{document}

\title{Covariant $c$-flation: a variational approach}

\author{Renato Costa} \email{Renato.Santos@uct.ac.za}
\affiliation{The Cosmology and Gravity Group, Department of Mathematics and Applied Mathematics, University of Cape Town, Private Bag, Rondebosch, 7700, South Africa}
\affiliation{Department of Physics, McGill University, Montr\'eal, QC, H3A 2T8, Canada}
\author{Rodrigo R. Cuzinatto} \email{cuzinatto@hep.physics.mcgill.ca}
\affiliation{Department of Physics, McGill University, Montr\'eal, QC, H3A 2T8, Canada}
\affiliation{Instituto de Ci\^encia e Tecnologia, Universidade Federal de Alfenas, Rodovia Jos\'e Aur\'elio Vilela, 11999, Cidade Universit\'aria, CEP 37715-400 Po\c cos de Caldas, MG, Brazil}
\author{Elisa G. M. Ferreira}\email{elisa.ferreira@mail.mcgill.ca}
\author{Guilherme Franzmann} \email{guilherme.franzmann@mail.mcgill.ca}
\affiliation{Department of Physics, McGill University, Montr\'eal, QC, H3A 2T8, Canada}

\date{\today}

\begin{abstract}
We develop an action principle to construct the dynamics that give rise to a minimal generalization of Einstein's equations, where the speed of light ($c$), the gravitational constant ($G$) and the cosmological constant ($\Lambda$) are allowed to vary. Our construction preserves general covariance of the theory, which yields a general dynamical constraint on $c$, $G$ and $\Lambda$. This action is general and can be applied to describe different cosmological solutions. We apply this formulation to the initial condition puzzles of the early universe and show that it generates a dynamical mechanism to 
%explain the initial conditions necessary to 
obtain the homogeneous and flat universe we observe today. We rewrite the conditions necessary to solve the horizon and flatness problems in this framework, which does not necessarily lead to an accelerated expansion as in inflation. Then, we show how the dynamics of the scalar field that represents $c$ or $G$ (and $\Lambda$) can be used to solve the problems of the early universe cosmology by means of different ways to c-inflate the horizon in the early universe. By taking $\Lambda = 0$, we show that the dynamics of the scalar field representing $c$ can be described once a potential is given. 
\end{abstract}

%\pacs{98.80.-k, 95.36.+x, 98.80.Es, 95.30.Sf, 98.80.Jk}

\maketitle

\section{Introduction}

The Standard Cosmological Model (SCM) is the basis for our understanding
of the universe. It provides an accurate description of the evolution
of an initial hot and dense universe expanding from times as early
as the nucleosynthesis until our current epoch. Despite all
of its successes, with the improvement of the observations and of
our understanding of the theory, we see that the SCM is incomplete
and  presents flaws. These flaws are particularly related to the
initial conditions of the universe, the apparent flatness of the universe today (\textit{flatness problem}), how regions that
are not causally connected are correlated (\textit{horizon
problem}), and the origin of all the structures in the universe
(\textit{origin of structures problem}). The initial conditions
of the SCM would have to be extremely fine tuned to solve all these
problems.

In the search for a mechanism that could dynamically
solve these puzzles, the theory of inflation \cite{Starobinsky:1980te,Guth:1980zm,Linde:1981mu,Albrecht:1982wi} was proposed. This was a period of accelerated expansion in the very early universe that
took place before the radiation-dominated decelerated expansion. Since
its proposal, inflation has become the current paradigm to describe
the evolution of the early universe not only by solving the SCM puzzles,
but also by providing a causal mechanism for the origin of the fluctuations
that seeded all the structures in the universe \cite{Starobinsky:1979ty,Mukhanov:1981xt}. 
This mechanism can be measured today with great precision from the
anisotropies in the Cosmic Microwave Background (CMB) radiation temperature \cite{Ade:2013zuv}, confirming
the prediction of an almost scale invariant and Gaussian power spectrum. Although
inflation achieves this incredible phenomenological success, it cannot
be fully confirmed by current data and requires further tests of its
predictions. Its microscopic and effective realizations also present
serious theoretical challenges (see \cite{Baumann:2009ds} for a review).

Therefore, it is interesting to investigate alternative models
to inflation that explain the current observations and solve
the inflationary early universe puzzles, but invoking different mechanisms. One class
of alternatives are bouncing models. In general, those models consist
of a universe that evolves from an initial phase of contraction followed,
after a bounce phase, by the usual SCM decelerated expansion. There
are many realizations of those models, such as the matter bounce \cite{Brandenberger:2012zb},
Ekpyrotic \cite{Khoury:2001wf,Buchbinder:2007ad}, and Pre-Big-Bang
\cite{Gasperini:1992em}.

An alternative way to overcome the SCM problems is to consider models
where the fundamental constants are allowed to vary. The idea of considering
variations of the fundamental constants is not new and can be dated
to the seminal works by Thomson and Taid \cite{Thomson_Taid}, and
Dirac \cite{Dirac}. 
The extensive searches for evidence of variation of the fundamental constants and studies of the potential observational consequences have been reviewed in Ref.~\cite{Uzan:2002vq}, for example. These variations are not completely excluded by current data, especially if they took place in the very early universe.

There have been many attempts to solve the initial condition problems of the SCM, with approaches exploring variation in different sets of fundamental constants. The initial
works by Moffat \cite{Moffat93}, Albretch and Magueijo \cite{AM} and Clayton and Moffat \cite{Clayton:1998hv} led this exploration by allowing the speed of light ($c$) to vary. These models are called Varying Speed of Light (VSL) and they solve the SCM problems by invoking a phase transition
in $c$, given by a fast change in its value, instead of a period
of inflationary expansion of the universe. Many variants
and different mechanisms for the variation of fundamental constants have been introduced so far,
as bimetric models \cite{Clayton:1998hv,Bassett:2000wj}, where
the coupling of matter and gravity are represented by different metrics;
Jordan-Brans-Dicke models \cite{B}, where variations of G come from
scalar-tensor theories; varying-$\alpha$ models \cite{BM,B}, where $\alpha$ is the fine
structure constant; and even VSL models from higher dimensional theories
\cite{higher_dim}.

In the majority of the models cited above, varying fundamental constants are introduced at the cost of either not conserving energy-momentum or violating Bianchi identity, thus sacrificing the covariance
of the theory. This is not true for all the models in the literature \cite{Avelino:1999is}. However,
many of these models only indicate the kinematics necessary for the varying constants to
solve the Big Bang puzzles, not providing a dynamical mechanism that explains this variation. This is one of the biggest criticisms encountered by this class of models. An action principle to describe the variation of such fundamental constants is key to the construction of this generalization of Einstein's gravity.
 
In this work, we confront this issue and develop an action principle that yields dynamics for the model from \cite{Gui}, where General Relativity (GR) is modified
by making the fundamental constants $c$, $G$ and $\Lambda$ depend on the spacetime coordinates. In this approach, the variation of the constants obey a general constraint coming from the Bianchi identity together with the standard local conservation laws. This formulation is general covariant (or diffeomorphism covariant) given that it preserves invariance under diffeomorphisms the same way GR does. The theory in our paper carries two extra propagating degrees of freedom; these can be reduced to one by neglecting the varying cosmological constant. The evolution of $c$ and $G$ can then be used to solve the SCM problems; our proposal requires conditions similar to the ones in some VSL models (i.e. violation the strong energy condition) or the ones needed to be imposed upon the SCM (e.g. inclusion of a $\Lambda(t)$). 

Given the general covariance of our model, we are able to promote the fundamental constants to scalar fields and develop a Lagrangian formulation of the theory through an invariant action. When developing an action for this minimal extension of GR  we strove to verify the internal consistency of the theory and
to determine the dynamics of the varying constants. Our generalized Einstein-Hilbert action can be used to study different cosmological solutions, such as inflation and even possibly dark energy\footnote{The later is a subject for
future work.}. Here we are interested in showing the different ways
in which our covariant framework allows us to inflate the cosmological horizon
using a variation of $c$, dubbed \textit{c-flation}, to solve the early universe
puzzles. One advantage of this model is that one does not need
to introduce an \textit{ad hoc} field to be responsible for inflating
the universe, as it happens with the inflaton. For different choices of the potential, the horizon
can be inflated by means of a period of accelerated expansion, or
by a sharp change in the value of $c$ or by means of a super/subluminal
expansion yielding a shorter inflationary period.

This paper is organized as follows: Section \ref{sec:model} describes
our setup, by defining our extension to Einstein's equations via the introduction of varying $c$, $G$ and $\Lambda$ and by writing the Friedmann equations
for a cosmological Friedmann-Robertson-Walker (FRW) background. Section \ref{sec:Puzzles} addresses how the particle horizon and the early universe puzzles
are described in this theory and presents the possible solutions that
our model can provide as well as the conditions that must be satisfied for
their validity. In Section \ref{5} we establish the Lagrangian formulation of this theory,
deriving the generalized Klein-Gordon equation that determines the
dynamics of the field representing $c$ (or $G$) and showing
how this can be used to solve the puzzles of the SCM. Finally, in
Section \ref{sec:conclusions} we present our discussions and conclusions. In Appendix \ref{sec:A}, we discuss how introducing a $\Lambda(t)$ can solve early universe problems and alleviate the cosmological constant problem; Appendix \ref{sec:B} shows how single field inflation can be conceptually understood within our varying fundamental constants framework.

\section{Covariant Varying ``Constants'' Model }\label{sec:model}
Recently, a covariant model of varying fundamental ``constants'' has been
revived in \cite{Gui} and some cosmological applications have been
discussed. In this section we review this model. 

\subsection{The General Constraint}

The idea is to perform a minimal change in GR by allowing the fundamental constants that appear in Einstein's equations and the cosmological constant to vary with respect to the spacetime coordinates, $G=G\left(x^{\alpha}\right)$,
$c=c\left(x^{\alpha}\right)$ and $\Lambda=\Lambda\left(x^{\alpha}\right)$, leading to:
\begin{equation}\label{eq:EinsteinEqLambda}
G_{\mu\nu}=\frac{8\pi G\left(x^{\alpha}\right)}{c\left(x^{\alpha}\right)^{4}}T_{\mu\nu}-\Lambda\left(x^{\alpha}\right) g_{\mu\nu}\,,
\end{equation}
where $G_{\mu\nu}=R_{\mu\nu}-\frac{1}{2}g_{\mu\nu}R$ is the Einstein tensor, $R_{\mu\nu}$ and $R$ are the Ricci tensor and scalar, respectively, and $T_{\mu\nu}$ is the energy-momentum tensor (EMT) of the theory being considered.

In order to build a consistent theory, Eq.~(\ref{eq:EinsteinEqLambda}) should satisfy the contracted Bianchi identities, which assumes a torsion-free connection and the metricity condition summarizing the underlying geometrical structure present in a metric manifold \cite{Carroll}, and local conservation laws:
\begin{equation}
\nabla^{\mu}G_{\mu\nu}=0\,, \qquad  \nabla^{\mu}T_{\mu\nu}=0\,.
\label{eq:nablaG}
\end{equation}
%Since the Einstein tensor is covariantly conserved which means it preserves the geometric structure based on a pseu-Riemanian geometry (Lorentzian geometry). So, we ma stress tensor cannot be obtained by variational differenti- ation of any local Lagrangian density based on a pseudo- Riemannian geometry.
This leads to the following constraint in the variation of such ``constants'':
\begin{equation}
\left[\frac{1}{G}\partial_{\mu}G-4\frac{1}{c}\partial_{\mu}c\right]\frac{8\pi G}{c^{4}}T^{\mu\nu}-\left(\partial_{\mu}\Lambda\right)g^{\mu\nu}=0\,.
\label{eq:GeneralConstraintEq}
\end{equation}
Following the nomenclature of \cite{Gui} we will call this result the \textit{general constraint} (GC). It is important to mention that although it is clear what is meant by $G$ and $\Lambda$ above, this is not the case for $c$. The terminology considered here is the one given in \cite{EU}, so that $c$ is the ``spacetime speed'', the speed that appears in the metric, which is considered equal to ``Einstein's speed'', the speed that appears in the coupling between matter content and geometry in Einstein's equations.

The theory constructed in this way has some important features that are worth stressing. First, we postulate \textit{ab initio} that the fundamental constants $c$, $G$ and $\Lambda$ are allowed to change. This means these ``constants'' also vary in any other theory containing them. This aspect will be explored in a future work, where we will study how electrodynamics changes when $c$ varies, for instance. Second, this theory preserves covariance, since it is invariant under diffeomorphisms the same way as it happens in GR. Einstein's tensor is covariantly preserved, maintaining the structure of a pseudo-Riemanian geometry. This allows us to obtain the covariantly conserved energy-momentum tensor from a Lagrangian density on a local Lorentzian geometry, as shown in the following sections. This procedure is not straightforward in the absence of such a geometrical structure. 

\subsection{Cosmological Background}

The SCM could be divided in two different scales: a homogeneous and isotropic one, which corresponds to cosmological scales, and an inhomogeneous one, which accounts for the structures observed in the universe. These scales can be described non-perturbatively and perturbatively, respectively. This can be appreciated on the CMB, since its fluctuations are of order $10^{-5}$ of its average temperature.

Therefore, if one disregards the fluctuations, the necessary description would be made exclusively out of a homogeneous and isotropic solution. As the fluctuations are turned on again, their description would be seen as fluctuations over the homogeneous and isotropic solution already established, as it is typically the case. It is important to overemphasize this point since then it becomes clear that the CMB only fixes a class among all the possible reference frames to be considered. 

Thus, the preferred coordinate system to be used is the one that assumes homogeneity and isotropy, which implies the \textit{ansatz} known as Friedmann-Robertson-Walker metric:
\begin{equation}
ds^{2}=-c^{2}(t)dt^{2}+a^{2}(t)\left[\frac{1}{1-kr^{2}}dr^{2}+r^{2}\left(d\theta^{2}+\sin^{2}\theta d\phi^{2}\right)\right]\,,\label{eq:FRW}
\end{equation}
where $k=-1,0,1$ determines the geometry of the space sector of the
manifold: hyperbolic, flat or spherical, respectively. Due to homogeneity and isotropy, the speed
of light is regarded as a function of time only, $c=c(t)$, which also implies $G=G(t)$ and $\Lambda=\Lambda(t)$. 

For a homogeneous and isotropic universe, the matter content can be properly described by a perfect fluid. Its EMT is defined to be:
\begin{equation}
T^{\mu\nu}=\frac{1}{c^2}(\varepsilon+p)U^\mu U^\nu +pg^{\mu\nu},
\end{equation}
where $\varepsilon$ is the energy density, $p$ is the pressure, and $U^\mu$ is the $4$-velocity satisfying $g_{\mu\nu}U^\mu U^\nu = -c^2$.

Generalized Friedmann equations can be derived from (\ref{eq:EinsteinEqLambda}) and (\ref{eq:FRW}),
\begin{eqnarray}
H^{2} (t)&=&\frac{8\pi G(t)}{3c^{2}(t)}\varepsilon(t)+\frac{\Lambda(t) c^{2}(t)}{3}-\frac{kc^{2}(t)}{a^{2}(t)}\,\label{eq:FirstFriedEq}\\
\frac{\ddot{a}(t)}{a(t)}&=&-\frac{4\pi G(t)}{3c^{2}(t)}\left[\varepsilon(t)+3p(t)\right]+\frac{\Lambda(t) c^{2}(t)}{3}+\frac{\dot{c}(t)}{c(t)}H(t),\label{eq:SecondFriedEq}
\end{eqnarray}
where $H=\dot{a}/a$ is the Hubble parameter. The second Friedmann
equation is modified with respect to the standard case including
a term proportional to $\dot{c}$. Combining both Friedmann equations one gets,
\begin{equation}
\dot{\varepsilon}+3\frac{\dot{a}}{a}\left(\varepsilon+p\right)=-\left[\left(\frac{\dot{G}}{G}-4\frac{\dot{c}}{c}\right)\varepsilon+\frac{c^{4}}{8\pi G}\dot{\Lambda}\right].\label{eq:GeneralContinuityEq}
\end{equation}
The right-hand side seems to violate the covariant conservation of $T^{\mu\nu}$, which reads $\nabla_{\mu}T^{\mu\nu}=\dot{\varepsilon}+3\frac{\dot{a}}{a}\left(\varepsilon+p\right)=0$.
However, in the context of background cosmology the GC in Eq.~(\ref{eq:GeneralConstraintEq})
reads:
\begin{equation}
\left(\frac{\dot{G}}{G}-4\frac{\dot{c}}{c}\right)\frac{8\pi G}{c^{4}}\varepsilon+\dot{\Lambda}=0,\label{eq:ConstraintEq}
\end{equation}
so that (\ref{eq:GeneralContinuityEq})
reduces to,
\begin{equation}
\frac{\dot{\varepsilon}}{\varepsilon}+3\frac{\dot{a}}{a}\left(1+\frac{p}{\varepsilon}\right)=0,\label{eq:ContinuityEq}
\end{equation}
showing that EMT conservation holds. This is not the
case in \cite{AM}, for instance, where the continuity equation is deliberately
violated. Incidentally, if $\dot{\Lambda}=0$ then (\ref{eq:ConstraintEq}) leads to:
\begin{equation}
G(t)=\frac{G_0}{c_0^4}c^4(t), \label{eq:simpleconstraint}
\end{equation}
which will be used in Section \ref{sec:Puzzles}.

\section{Early Universe Puzzles}\label{sec:Puzzles}

The initial success of inflationary models was basically due to its
background cosmology, which was able to explain why the universe looks
so homogeneous and isotropic, and also spatially flat on large scales.
Therefore, the very first test one can demand out of the current proposal
is to verify in which conditions those old problems could be solved.
In order to do so, we need first to define the particle horizon, since
now the speed of light is not constant anymore.

\subsection{Particle Horizon}

We start off considering our metric \textit{ansatz} (\ref{eq:FRW}),
\begin{equation}
ds^{2}=-c^{2}\left(t\right)dt^{2}+a^{2}\left(t\right)d\chi^{2},
\end{equation}
where $\chi$ are comoving coordinates.

Before proceeding any further, it is important to justify why massless particles would still follow null geodesics even when the limiting speed of the spacetime is changing. In order to do that, note that in GR we still have Lorentz covariance locally. Given that, one can define local Lorentz transformations depending explicitly on some constant spacetime speed $c_{1}$; this is necessary to keep the line element invariant under inertial transformations. It then follows from Special Relativity (SR) that the
energy and momentum of a particle are given by,
\begin{equation}
E:=\gamma mc_{1}^{2}, \qquad{}\vec{p}:=\gamma m\vec{v},
\end{equation}
where $\gamma^{-1}:=\sqrt{1-v^{2}/c_{1}^{2}}$ and $v$ is the velocity
of the particle. These two quantities can be combined into a single
equation,
\begin{equation}
E^{2}=m^{2}c_{1}^{4}+p^{2}c_{1}^{2}.\label{eq:EinsteinEq}
\end{equation}
Thus, it is easy to see that $v=c_1$ for any massless particle, which implies a null line element to these particles.

This notion needs to be generalized for the case where the speed of light is changing. The way to proceed here is the following: after we have foliated the spacetime using a time coordinate, let us focus on a particular leaf labelled by $t_{*}$, and then consider all the leaves in the
interval $t_{*}\pm\delta t$. For these leaves, the local geometric
structure can be thought as,
\begin{equation}
ds^{2}=-c_{*}^{2}dt^{2}+d\vec{x}^{2},
\end{equation}
where $c(t_{*})=c_{*}=\text{const.}$ and $\vec{x}$ are the physical coordinates. This also defines a SR theory locally with a limiting speed given by $c_{*}$, with which the massless
particles will be traveling; these particles will consequently move along a null geodesic. This processes can be repeated over and over again for all the leaves in our foliated spacetime leading to the conclusion that massless particles follow $ds^2=0$ even for a spacetime admiting $c=c(t)$.

We are now in position to define the particle horizon. Given our \textit{ansatz} (\ref{eq:FRW}), a null line element implies:
\begin{equation}
d\chi=\pm\frac{c\left(t\right)}{a\left(t\right)}dt,
\end{equation}
where we should think of $\chi$ as just the radial direction since the
spacetime is isotropic. Thus, the particle horizon is defined as:
\begin{equation}
d_{p}\left(t\right)=a\left(t\right)\int_{t_{i}}^{t}\frac{c\left(t'\right)}{a\left(t'\right)}dt'\,. \label{eq:ParticleHorizon}
\end{equation}
It can be re-written as:
\begin{equation}
d_{p}\left(t\right)=a\left(t\right)\int_{a_{i}}^{ a(t)}\left[\left(aH\right)^{-1}c\right]d\ln a\,.\label{eq:ParticleHorizon1}
\end{equation}
Note that if $c(t)=\mathrm{const.}$ this reduces to the usual GR expression \cite{Baumann:2009ds}.
The particle horizon tells us the maximum distance massless particles
can travel between $t_{i}$ and $t>t_{i}$. The comoving particle horizon is defined by $d_{p}(t)/a(t)$.

Note that if we assume a constant equation of state parameter $\omega= p/\varepsilon= \text{const}.$ for a perfect fluid and consider the continuity equation (\ref{eq:ContinuityEq}), we have that $\varepsilon\varpropto a^{-3\left(1+\omega\right)}$. Substituting this result into Eq.~(\ref{eq:FirstFriedEq}) and taking $k=\Lambda=0$, 
\begin{equation}\label{eq:resultSCM}
\left(aH\right)^{-1}c=c_0 H_{0}^{-1}a^{\frac{1}{2}\left(1+3\omega\right)},
\end{equation}
where we have used the constraint (\ref{eq:simpleconstraint}). We conclude that the comoving particle horizon for a $c=c(t)$ universe dominated by a fluid with a constant equation of state is:
\begin{equation} \label{eq:co_part_hor}
\frac{d_{p}\left(t\right)}{a\left(t\right)}=\frac{2 c_0 H_{0}^{-1}}{1+3\omega}\left.a^{\frac{1}{2}\left(1+3\omega\right)}\right|_{a_{i}}^{a}\,,
\end{equation}
which is the same expression one would recover in GR. Hence, we also see here that the contribution associated to $a_{i}\rightarrow0$ is negligible for $1+3\omega>0.$ This is intrinsically associated
to the fact that the integral kernel in (\ref{eq:ParticleHorizon1}) grows with time for $\omega>-1/3$,
\begin{equation}
\frac{d}{dt}\left[\left(aH\right)^{-1}c\right]>0\,.
\end{equation}
This means that the comoving Hubble radius increases as the universe expands. Therefore, the comoving particle horizon will be finite at a given time.

We have just seen that even after considering a varying speed of light, we do not get a particle horizon different from the one in the SCM. (This stems from the fact we satisfy the perfect fluid continuity equation and impose the general constraint). The lack of novelty in Eq.~(\ref{eq:co_part_hor}) already hints a non-trivial solution for the horizon and flatness problems within our framework. They are our next subjects.

\subsection{The Horizon Problem}

The CMB tells us that the universe looked very isotropic and homogeneous
after the decoupling of photons from the primordial plasma about $380,000$ years after the Big Bang. However, as we remarked above, we have seen that the particle
horizon for a radiation ($\omega=1/3$)- or matter ($\omega=0$)-
dominated universes is finite between the initial time, $t_{i}=0$,
and the time the CMB decoupled. This becomes a problem since it means
most of those photons were not in causal contact, a necessary condition
for them to thermalize producing the CMB with the properties
we observe today. This defines the \textit{horizon problem}.

In order to solve this we demand that the
particle horizon increases towards the past, so that we do not have
a negligible contribution coming from its lower bound. For a background
dominated by a perfect fluid, this is intrinsically related to demand\gui{ing} $1+3\omega<0$. More generally, there should be a phase in which,
\begin{equation}
\frac{d}{dt}\left[\left(aH\right)^{-1}c\right]=\frac{d}{dt}\left(\frac{c}{\dot{a}}\right)<0.\label{equation: solvingpuzzles}
\end{equation}
This phase will be called \textit{c-flation}. Of course, in GR the
speed of light is considered constant and this phase is known as inflation.
Actually, we show in Appendix \ref{sec:B} that standard single field inflation can be seen as a subset of the broader $c$-flation scenario, which provides a conceptual shift with respect to the inflaton.

The $c$-flation phase only solves the early universe puzzles if it lasts enough time. Accordingly, we demand that our current observable universe should fit in the comoving Hubble
radius at the beginning of $c$-flation, meaning:
\begin{equation}
\left(a_{0}H_{0}\right)^{-1}c_{0}<\left(a_{I}H_{I}\right)^{-1}c_{I},\label{eq:Cond1}
\end{equation}
where sub-index ``$0$'' denotes our current time while
``I'' denotes the time when $c$-flation started. Note
that if we assume that the universe has been dominated by radiation
since the end of $c$-flation (ignoring recent epochs associated to matter-
and dark energy-dominations), we can use (\ref{eq:resultSCM}),
\[
\left(aH\right)^{-1}c\varpropto\left.a^{\frac{1}{2}\left(1+3\omega\right)}\right|_{\omega=1/3}=a,
\]
to conclude:
\begin{equation}
\frac{a_{0}H_{0}}{c_{0}}\frac{c_{E}}{a_{E}H_{E}}\sim\frac{a_{e}}{a_{0}}\sim\frac{T_{0}}{T_{E}}\sim10^{-28},
\end{equation}
where label ``E'' denotes the time when $c$-flation ended. We have assumed $T_{E}\sim10^{15}$ GeV and $T_{0}\sim\,10^{-4}$eV.
Plugging the above relation back into (\ref{eq:Cond1}), we get:
\begin{equation} \label{eq:HorizonProb}
\frac{c_{I}}{a_{I}H_{I}}\frac{a_{E}H_{E}}{c_{E}}>10^{28}.
\end{equation}
A more standard notation in terms of \textit{e-folds} reads\footnote{An \textit{e-fold} is defined by $N=\ln a$. It measures the amount of elapsed Hubble times through $dN = H dt$.}:
\begin{equation}
\ln\frac{a_{E}}{a_{I}}+\ln\frac{H_{E}}{H_{I}}-\ln\frac{c_{E}}{c_{I}} \gtrsim 64.\label{eq:Cond2}
\end{equation}
This is the condition that the $c$-flationary phase has to satisfy in order to solve the horizon problem. For inflation, we have $H\sim const.$ and $c=\mathrm{const.}$, so that we recover the usual result, $\ln\left(a_{E}/a_{I}\right)>64,$ corresponding
to an expansion of $\sim64$ \textit{e-folds}. This notion will be generalized below when we discuss slow-roll in $c$-flation, where the ratio $H/c$ will be taken constant, instead of just $H$ and $c$ by themselves.

%%%%% SUB-SECTION %%%%%

\subsection{The Flatness Problem \label{sec:Flatness}}

Let us start by defining the cosmological parameter, $\Omega$, as:
\begin{equation}
\Omega:=\frac{\varepsilon(t)}{\varepsilon_c(t)},\label{eq:CosmoParameter}
\end{equation}
where the critical energy density is defined as $\varepsilon_{c}(t):=3H^{2}c^{2}/8\pi G$.
Thus, one can show that:
\begin{equation} \label{eq:Omega(k)}
\Omega-1=\frac{c^{2}k}{\left(aH\right)^{2}},
\end{equation}
after taking $\Lambda=0$ in (\ref{eq:FirstFriedEq}). Writing Eq.~(\ref{eq:Omega(k)}) at the time of $c$-flation onset and today, it follows:
\begin{equation} \label{eq:FlatnessProb}
\Omega_{I}-1=\left(\Omega_{0}-1\right)\left[\frac{c_{I}}{\left(aH\right)_{I}}\frac{\left(aH\right)_{0}}{c_{0}}\right]^{2}\leq10^{-56},
\end{equation}
given that we observe $\Omega_{0}\sim1$ \cite{Ade:2013zuv}. Constraint (\ref{eq:FlatnessProb}) is fulfilled only if $\Omega_{I}$ was extremely close to unity. This fine tuning problem is known as \textit{flatness problem}.

It can be rephrased in the following way:
\begin{eqnarray}
\Omega_{0} & \sim & 1+10^{56}\left(\Omega_{I}-1\right)\left[\frac{\left(aH\right)_{I}}{c_{I}}\frac{c_{E}}{\left(aH\right)_{E}}\right]^{2},
\end{eqnarray}
so that in order to have $\Omega_{I}-1\sim\mathcal{O}\left(1\right)$
we do need:
\begin{equation}
\frac{\left(aH\right)_{I}}{c_{I}}\frac{c_{E}}{\left(aH\right)_{E}}<10^{-28},
\end{equation}
which is equivalent to the condition we derived above for the horizon
problem, Eq.~(\ref{eq:HorizonProb}).

%%%%% SUB-SECTION %%%%%

\subsection{Solving the SCM puzzles \label{sec:SolvingPuzzles}}

As stated above, in order to solve the horizon and flatness problems we need  an early-universe phase during which the comoving particle horizon is shrinking,
\begin{equation} \label{eq:shrink}
\frac{d}{dt}\left[\left(aH\right)^{-1}c\right]=\frac{d}{dt}\left(\frac{c}{\dot{a}}\right)<0\,.
\end{equation}
and it has to shrink the amount consistent with:
\begin{equation} 
\ln\frac{a_{E}}{a_{I}}+\ln\frac{H_{E}}{H_{I}}-\ln\frac{c_{E}}{c_{I}}>64.\label{eq:Cond2}
\end{equation}
In standard inflation (with no varying constants), this condition implies that the universe is accelerating, $\ddot{a} > 0$. It also implies that the strong energy condition, $w <-1/3$, is violated and the Hubble parameter is varying slowly, \textit{i.e.} $\epsilon_{sr} = - \dot{H}/H^2 < 1$, where the label \textit{``sr''} stands for \textit{slow-roll}. This is not true in our case. Here, there are other mechanisms that can lead to the shrinking of the particle horizon, and to $c$-flation.

%\sout{First, we can see that \rod{\sout{it} $c$-flation} does not necessarily imply acceleration}\rod{\sout{, since the condition now is}. 
In effect, Eq.~(\ref{eq:shrink}) demands:
\begin{equation} \label{eq:Cond-cflation}
\ddot{a} > \dot{a} \frac{\dot{c}}{c} \,.
\end{equation}
Clearly, inflation is recovered when $\dot{c}=0$, \textit{i.e.} for a constant $c$. More generally, according to our picture $c$ might as well vary provided condition (\ref{eq:Cond-cflation}) is satisfied. For example, if this variation is positive and $\dot{a} > 0$, we will have an accelerated expansion, even with $c$ changing. Another interesting possibility is that this model can even accommodate an evolution where universe that is decelerating, $\ddot{a} <0$: one case comes to be if the variation of $c$ is decreasing, $\dot{c} < 0$, while $\dot{a} > 0$; another one if the variation of $c$ is increasing but we are in a contracting universe, with $\dot{a} <0$. All of those are equivalent ways of solving the horizon and flatness problems. A summary of the possibilities is given in Table \ref{tab:table}. Therefore, not only can one study accelerated expansion, but also decelerated one in this framework, which might lead to interesting applications to bouncing models.

%\rod{Clearly, inflation is recovered when $\dot{c}=0$, \textit{i.e.} for a constant $c$. More generally, according to our picture $c$ might as well vary provided condition (\ref{eq:Cond-cflation}) is satisfied. For example, there will be an accelerated expansion if both the speed of light is increasing ($\dot{c}>0$) and the universe is expanding ($\dot{a}>0$). Surprisingly, an accelerated contracting universe ($\dot{a}<0$) is able to solve horizon and flatness problems as long as the speed of light decreases ($\dot{c}<0$) simultaneously. This last example, shows our framework applicability might reach out to bouncing models. Other possibilities even accommodate decelerated scenarios: in fact, Eq.~(\ref{eq:Cond-cflation}) remains valid if $\dot{a}>0$, $\dot{c}<0$ and $-|\dot{a}\dot{c}/c|<\ddot{a}<0$}

%\gui{(I don't follow the last inequality. If $\dot{c}$ is negative and $\dot{a}$ is positive, then clearly $\ddot{a}$ can be either positive or negative, since the RHS of the inequality will be negative. Or the idea was just to emphasize a particular case?)}\notere{(I agree with Guilherme. Rodrigo, what did you have in mind here?)} \rod{$\leftarrow$ \textbf{Rodrigo:} I started the sentence as ``Other possibilities accommodate decelerated scenarios:''. This implies $\ddot{a}<0$. However, I do not oppose writing simply $-|\dot{a}\dot{c}/c|<\ddot{a}$. By the way, I do not oppose kepping the original text either, if you prefer.}. Table \ref{tab:table} gives a summary of the possible configurations resolving flatness and horizon problems within $c$-flation.}

\begin{table}[H]
\begin{tabular}{|c|c|c|}
 \hline
 & $\,\,\dot{a}<0 \text{ (contracting)}\,\,$ & $\,\,\dot{a}>0 \text{ (expanding)}\,\,$\tabularnewline
\hline
$\dot{c}>0 \text{ (subluminal)}$ & $\ddot{a}>-\left|\dot{a}\dot{c}/c\right|$& $\ddot{a}>\left|\dot{a}\dot{c}/c\right|$ \tabularnewline
\hline
$\dot{c}<0 \text{ (superluminal)}$ & $\ddot{a}>\left|\dot{a}\dot{c}/c\right|$ & $\ddot{a}>-\left|\dot{a}\dot{c}/c\right|$\tabularnewline
\hline
\end{tabular}
\centering \caption{Summary of possibilities that inflate the particle horizon. Sub- and super-luminal are defined in relation to $c(t)$ being smaller or bigger than the standard value of the speed of light $c_0$, respectively. It is assumed that $c$ evolves monotonically in time.}
\label{tab:table}
\end{table}

Let us see qualitatively how the evolution of $c$ affects the conditions
we need to fulfill. Consider the following expansion,
\begin{equation}
c\left(t\right)\simeq c_{0}+\dot{c}\left(t_{E}\right)\left(t-t_{E}\right).
\end{equation}
 Time value $t_{E}$ indicates the end of a dynamical $c$. Note that if the derivative is positive, it means $c$ has been
increasing towards $c_{0,}$ characterizing a subluminal phase; if $\dot{c}\left(t_{E}\right) < 0$, then a superluminal phase is described. In particular, during slow-roll $c$-flation,
we have:
\begin{equation}
\frac{H}{c}  =  \text{const.}=\frac{H_{E}}{c_{0}},
\end{equation}
as will be discussed later. Then, 
\begin{eqnarray}
\ln\frac{a_{E}}{a_{I}} & = & \frac{H_{E}}{c_{0}}\int dtc\left(t\right)\nonumber \\
 & \simeq & H_{E}\Delta t-\frac{1}{2}\frac{H_{E}}{c_{0}}\dot{c}\left(t_{E}\right)\Delta t^{2}\simeq60. \label{eq:sub-super-luminal}
\end{eqnarray}
Eq.~(\ref{eq:sub-super-luminal}) is an estimate for the $c$-flation \textit{e-fold} number;  in the case of single field inflation, the period of acceleration must last around $60$ \textit{e-folds} in order to solve the SCM problems. Eq.~(\ref{eq:sub-super-luminal}) then tells us a subluminal $c$-flation requires larger $H_E \Delta t$ values to cope with the negative sign on the second line; this means the theory demands a longer dynamical phase for the speed of light. Conversely, superluminal $c$-flation permits a shorter $c=c(t)$ period.

\vspace{0.5cm}

Unlike what happens in the slow-rolling inflationary case, $c$-flation shrinking comoving particle horizon does not imply that the Hubble parameter is varying slowly. In our case, 
\begin{equation}
\epsilon_{sr} + \frac{\dot{c}}{cH} < 1\,,
\label{new_hubble_change_cond}
\end{equation}
where the standard slow-roll parameter $\epsilon_{sr}$ was defined above, $ \epsilon_{sr}=-\dot{H}/H^2 $. Here, we have different ways to satisfy this condition involving changes in $c$ during a Hubble time. At this point we %\rod{\sout{can define a new slow-roll parameter that measures the change in the Hubble radius in the varying fundamental constants framework.  For that, we can}} 
define a new dimensionless generalized slow-roll parameter respecting condition (\ref{new_hubble_change_cond}):
\begin{equation} \label{eq:inflatinghorizon}
\epsilon_c = - \frac{1}{c} \frac{(H/c)^{\cdot}}{(H/c)^2} = - \frac{d \ln (H/c)}{dN} < 1\,,
\end{equation}
where $dN\equiv Hdt$. Inflation demands that the condition $\epsilon_{sr}<1$ be valid for an amount of time consistent with the smallness of the second slow-roll parameter $\eta_{sr}$ \cite{Baumann:2009ds}. In analogy with the standard inflationary theory, we define generalized second slow-roll parameter:
\begin{equation}
\eta_c = \frac{d \ln \epsilon_c}{dN} = \frac{1}{H}\frac{\dot{\epsilon_c}}{\epsilon_c}\,, \label{eq:etac}
\end{equation}
which measures the change in $\epsilon_c$ in a Hubble time, and we require its absolute value to be smaller than one during $c$-flation, $|\eta_c|<1$. However, one needs to see that this condition is not necessary, and it is not general in our model, since it does not encompasses the case of a phase transition in $c$, as we will see in the following sections.
One comment is in order here. We were able to describe $c$-flation in terms of generalized slow-roll parameters that are equivalent to the condition of shrinking particle horizon. However, one must remember that this is different than standard inflation, since effects coming only from the variation of $c$ can be responsible for solving the early universe puzzles.
%(\notere{I kept the original text here and slightly changed the beginning of the next paragraph.})
%\elisa{!!!Eu prefiro o texto que estava antes.!!!}
%\rod{(The condition of small slow-roll parameter values could be relaxed in the particular case of a phase transition in $c$. This phase transition is a feature of VSL models; it is allowed within our framework through changes in $\dot{c}/c$ -- see Eq.~\ref{eq:Cond-cflation}.)} 

%%%%%%% (mix both?)Pq "shrinking..." foi dito no parágrafo anterior (se mantiver o antigo)
%\rod{\sout{We saw that, different than the inflationary case, here condition (\ref{equation: solvingpuzzles}) does not necessarily imply acceleration with slow-roll.}} \rod{
The description of $c$-flation in terms of generalized slow-roll parameters is equivalent to condition (\ref{equation: solvingpuzzles}). 
%of shrinking particle horizon.} 
It implies that we need to violate the strong energy condition with $\omega < -1/3$, as it is the case for standard inflation. In the presence of a varying cosmological constant ($\Lambda\neq0$),  Eq.~(\ref{equation: solvingpuzzles}) is equivalent to:
\begin{equation}
3\varepsilon\left(\omega+\frac{1}{3}\right)-\frac{c^{4}}{4\pi G}\Lambda<0\,.
\end{equation}
It is no surprise that a positive cosmological constant-like term could inflate our particle horizon. We explore this possibility in Appendix \ref{sec:A}.

 We conclude this section by recalling that the $c$-flation framework encompasses different ways to solve SCM puzzles. It contains other approaches found in the literature as particular cases or similar mechanisms. Standard inflation is itself an example, although it usually does not incorporate any varying constants\footnote{However, see Appendix \ref{sec:B} for an interesting perspective.}.
Contracting solutions can also be used to solve the initial condition puzzles within $c$-flation proposal. 
%\rod{\sout{Variations of $\Lambda$ that also cause a violation of the strong energy condition constitute other models that are presented in the literature (see \cite{Avelino:1999is}, for example), and are also included here, as it can be seen in Appendix \ref{sec:A}. The VSL models invoked a different mechanism to solve those problems, namely a phase transition in $c$. This is also allowed here, since we can see from above that changes in $\dot{c}/c$ might solve the puzzles. We show this in more detail in the next section.}} 
Models with varying $\Lambda$ are also included by $c$-flation mechanism; together with the strong energy condition violation, they are able to cause the shrinking of particle horizon. This conclusion resonates with the literature -- see e.g. \cite{Avelino:1999is}. For all its embracing power, $c$-flation framework deserves a careful analysis in terms of fundamental field dynamics as opposed to \textit{ad hoc} hypotheses. This brings us to some motivations for the next section.

\vspace{0.5cm}

It is important to emphasize that we have learned considerably from the inflationary paradigm regarding
how we can emulate the conditions for a shrinking horizon from a scalar field. Notwithstanding, inflation relies on the introduction
of an \textit{ad hoc} scalar field whose fundamental theory is yet to be found and properly motivated. Our $c$-flation framework bears in principle three new scalar fields, one associated to the speed of light, another to Newton's coupling, and a third one related to $\Lambda$. These fields are fundamentally constrained by the GC, leaving only two independent degrees of freedom. In fact, for the most part, we will be disregarding the cosmological constant-like term; this restriction actually leaves us with only one degree of freedom, $c$ or $G$ -- see Eq.~(\ref{eq:simpleconstraint}). In the following section, this independent degree of freedom will be promoted to a scalar field.
%\gui{. \sout{; by  defining a consistent set of equations of motion coming from a well-defined action,}} \notere{\st{we shall propose a powerful framework that can establish a new paradigm for varying fundamental constants in a sense similar to what the inflationary paradigm represents today.}}

%%%%%%% SECTION %%%%%%%

\section{Covariant C-flation}\label{5}

A common criticism to most VSL proposals is based on the fact that the variation of the constants was necessarily \textit{ad hoc} \cite{Ellis:2007ah}. 
%We have shown that within our framework, we would not be able to solve the horizon and flatness problem without $\Lambda$ \gui{(wtf?)}. Instead, we have shown that we can introduce exotic matter that violates the strong energy condition, which solves these puzzles. This is analogous to standard GR.
Our goal in this section is to fill in this gap by promoteing the fundamental constants to fields and consequently introducing an action that is simultaneously consistent with Einstein's equations and the GC -- Eq.~(\ref{eq:GeneralConstraintEq}). %\rod{We claim this is a significant gain with respect to the previous approaches since the extra scalar field which is meaningfully introduced and which leads to a time dependence of the fundamental constants of our four space-time dimensional world could be viewed as having an origin in some higher dimensional physics\footnote{\rod{We are grateful to the referee for this insightful comment.}}. }
%\elisa{!!! Eu acho que não deveríamos inserir esse comentário comecando em "We claim..."!!! Se quiserem inserir acho que no máximo na conclusão e sem citar o referee num footnote.}
Moreover, our approach will differ from the standard inflation paradigm in the sense that our fields will not be introduced in an \textit{ad hoc} fashion.

We start by writing down a generalized Einstein-Hilbert action (gEH) defined as\footnote{Note that our gEH action is different from the ones in Refs. \cite{Moffat06,Moffat16}.}: 
\begin{equation}\label{gEH}
S_{gEH}=\frac{1}{16\pi}\int d^{4}x\sqrt{-g(\Phi)}\bigg\{\frac{\Phi}{\Psi}(R(\Phi)-2\Lambda)+\lambda^{\mu}\bigg[\bigg(\frac{1}{\Psi}\partial^{\nu}\Psi-\frac{4}{\Phi^{1/4}}\partial^{\nu}\Phi^{1/4}\bigg)\frac{8\pi\Psi}{\Phi}T_{\mu\nu}-g_{\mu\nu}\partial^{\nu}\Lambda\bigg]\bigg\},
\end{equation}
where $c^{4}(x^\alpha)\equiv\Phi(x^\alpha)$,
$G(x^\alpha)\equiv\Psi(x^\alpha)$ and $\Lambda=\Lambda(x^\alpha)$ are the scalar fields related to the speed of light, gravitational coupling and cosmological term respectively. We have introduced a Lagrange multiplier, $\lambda^\mu$, that imposes the GC on shell in the action. In fact, varying the action with respect to $\lambda^\mu$ gives
\begin{equation}
\bigg(\frac{1}{\Psi}\partial^{\nu}\Psi-\frac{4}{\Phi^{1/4}}\partial^{\nu}\Phi^{1/4}\bigg)\frac{8\pi\Psi}{\Phi}T_{\mu\nu}-g_{\mu\nu}\partial^{\nu}\Lambda=0\,.\label{GC0}
\end{equation}

Note that the presence of the energy-momentum tensor inside the action may lead to redundancies in the definition of the very same tensor; however, see \cite{Ayuso:2014jda}, \cite{Sami:2002se} and \cite{Joyce:2014kja} for examples of well-defined theories. Additionally, we remark that the full constraint in Eq.~(\ref{eq:GeneralConstraintEq}) is non-holonomic: one cannot solve it and use the solution to eliminate degrees of freedom in the action. These two points are not important for our subsequent calculations, as we will only consider the case where the ``cosmological constant'' is set to zero. In this case (i) the constraint becomes holonomic and we can use it to eliminate degrees of freedom of the action, (ii) the energy-momentum tensor does not play any role in the variational procedure after step (i).

To allow the fundamental constants to vary in spacetime is equivalent to introducing three new scalar fields. Therefore, it is necessary to add dynamics for them as well. This can be done by defining the matter action as:
\begin{equation}
S_{m}=S_{sm}+S_{vc}\,,\label{matter}
\end{equation}
where $S_{sm}$ is the action associated to the standard matter content and,
\begin{equation}
S_{vc}=\frac{1}{16\pi}\int d^{4}x\sqrt{-g}\bigg\{-\kappa_{1}\frac{1}{2\Phi\Psi}\nabla_{\mu}\Phi\nabla^{\mu}\Phi-\kappa_{2}\frac{\Phi}{2\Psi^{3}}\nabla_{\alpha}\Psi\nabla^{\alpha}\Psi-\kappa_{3}\frac{\Phi}{2\Psi\Lambda^{2}}\nabla_{\nu}\Lambda\nabla^{\nu}\Lambda-V(\Phi,\Psi,\Lambda)\bigg\} \label{eq:action}\,,
\end{equation}
is the action giving dynamics to the varying ``constants''. In Eq.~(\ref{eq:action}) $\kappa_{1}$, $\kappa_{2}$ and $\kappa_{3}$ are dimensionless parameters and $V(\Phi,\Psi,\Lambda)$ is the potential related to these fields. The nontrivial couplings in the kinetic terms above are introduced in order to assure the correct dimensions.

Let us emphasize that we have more degrees of freedom than in standard GR. There are three scalar fields and one constraint equation; this adds two new degrees of freedom to standard GR in a consistent covariant way. Although the steps to derive the equations of motion for all the fields are clear, they are are not so simple to follow in the general case. The easier path is taking $\Lambda=0$; it allows us to solve the GC and this is what we will do next.

%%%%% SUB-SECTION %%%%%

\subsection{A homogeneous and isotropic universe without a cosmological ``constant''}

If we derive action (\ref{gEH}) with respect to the Lagrange multiplier, we get
\begin{equation} 
\bigg(\frac{1}{\Psi}\partial^{\nu}\Psi-\frac{4}{\Phi^{1/4}}\partial^{\nu}\Phi^{1/4}\bigg)\frac{8\pi\Psi}{\Phi}T_{\mu\nu}=0.\label{eq:LagMultiEq}
\end{equation}
Given that we are not interested in vacuum solutions,  Eq.~(\ref{eq:LagMultiEq}) gives:
\begin{equation}
\Psi(x^\alpha)=\frac{G_{0}}{c_{0}^{4}}\Phi(x^\alpha)\label{sol1},
\end{equation}
where $G_{0}$ and $c_{0}$ are the currently constant values of Newton's constant and the speed of light. We assume  $\Phi(t_0)=c_{0}^{4}$ with $t_0$ as the age of the Universe in order to
recover the GR limit. Using (\ref{sol1}), we reduce (\ref{eq:action}) to\footnote{ The kinetic term of our model is similar to the one in Jordan-Brans-Dicke action (see e.g. \cite{Uzan:2002vq}),
\eqn{\label{Jordan}
S = \int d^4x \sqrt{-g} \phi^\eta \left[R - \xi \frac{\nabla^\mu \phi \nabla_\mu \phi}{\phi^2} - \frac{\phi}{2}F^2\right].
}
Notice, however, that our model in Eq.~(\ref{eq:Svc(Phi)}) is fundamentally different 
from (\ref{Jordan}) even in the case where $\eta = 0$ and $F =0$. In fact, Jordan-Brans-Dick theory lacks a dynamical constraint between $c$ and $G$. Moreover, the speed of light for that model is constant, unlike ours. }
\begin{equation} \label{eq:Svc(Phi)}
S_{vc}=\frac{1}{16\pi}\int d^{4}x\sqrt{-g(\Phi)}\left(-\frac{c_{0}^{4}(\kappa_{1}+\kappa_{2})}{G_{0}}\frac{1}{2\Phi^{2}}\nabla_{\mu}\Phi\nabla^{\mu}\Phi-V(\Phi)\right).
\end{equation}
The field redefinition
\begin{equation} \label{eq:fieldredefinition}
\Phi\rightarrow\phi\equiv\sqrt{(\kappa_{1}+\kappa_{2})c_{0}^{4}/16\pi G_{0}}\ln(\Phi/c_{0}^{4}) \,,
\end{equation}
 casts (\ref{eq:Svc(Phi)}) into the form:
\begin{equation}
S_{vc}=\int d^{4}x\sqrt{-g(e^{\phi/\phi_{0}})}\left(-\frac{1}{2}\nabla_{\mu}\phi\nabla^{\mu}\phi-V(\phi)\right)\,,
\end{equation}
where $\phi_{0}=\sqrt{(\kappa_{1}+\kappa_{2})c_{0}^{4}/16\pi G_{0}}$.
Therefore, the total action becomes,
\begin{equation}
S=\int d^{4}x\sqrt{-g(e^{\phi/\phi_{0}})}\left[\frac{c_{0}^{4}}{16\pi G_{0}}R(e^{\phi/\phi_{0}})-\frac{1}{2}\nabla_{\mu}\phi\nabla^{\mu}\phi-V(\phi)\right]+S_{sm}\,.
\end{equation}
Note that the above action seems to be describing a standard scalar field in a curved background, but that is not the case. This action specifically describes the scalar field related to the speed of light (or, equivalently, to Newton's coupling). The FRW line element displays $c$; consequently it depends on the field $\phi$. In fact,
\eqn{\label{mgEH1} ds^2 = -e^{\phi/2\phi_0}c_0^2dt^2 +
a^2(t) d\vec{x}^2. }
where
\begin{equation}
\frac{\phi}{\phi_{0}}=4\ln\left(\frac{c}{c_{0}}\right),
\end{equation}
has a clear interpretation: scalar field $\phi$ is  born from a varying speed of light. Conversely, the inflationary scenario offers no explanation whatsoever for the nature of the inflaton field. The next step is to derive the equation of motion for $c$-flation field $\phi(t)$ and to check whether it can solve SCM puzzles within our variational approach.

%%%%% SUB-SECTION %%%%%

\subsection{Equations of Motion}

We have already computed Friedmann equations for a perfect fluid above, Eqs.~(\ref{eq:FirstFriedEq}) and (\ref{eq:SecondFriedEq}). Thus, we only need to calculate the energy density and pressure associated with the scalar field.

The energy-momentum tensor related to $S_{vc}$ can be computed using the definition
\eqn{ T^{vc}_{\mu\nu}  \equiv - \frac{2}{\sqrt{-g}}
\frac{\delta S_{vc}}{\delta g^{\mu\nu}}, }
which gives
\eqn{ T^{00}_{vc} &=& \frac{1}{c_0^2 e^{\phi/2\phi_0}}\left[\frac{1}{c_0^2 e^{\phi/2\phi_0}}
\frac{\dot{\phi}^2 }{2} + V(\phi)\right]
=  -g^{00}\varepsilon_{\phi},
\\
 T^{ij}_{vc} &=&- \frac{1}{a^2} \left[ \frac{1}{c_0^2 e^{\phi/2\phi_0}}
\frac{\dot{\phi}^2}{2} -V(\phi) \right]
= -g^{ij} p_\phi, }
so that,
\eqn{
\varepsilon_\phi=
\frac{1}{c_0^2 e^{\phi/2\phi_0}} \frac{\dot{\phi}^2
}{2} + V(\phi), \qquad{}p_\phi = \frac{1}{c_0^2e^{\phi/2\phi_0}}
\frac{\dot{\phi}^2}{2} -V(\phi).  \label{energy_presure}
}

The first generalized Friedmann equation (\ref{eq:FirstFriedEq}) is then written as:
\eqn{\label{gfe1}
H^2 = \left( \frac{\dot{a}}{a}\right)^2 = \left[ \frac{8\pi G_0 }{3c_0^2}\varepsilon - \frac{kc_0^2}{a^2} \right] e^{\phi/2\phi_0}\,\quad \quad (\Lambda=0),
}
where $\varepsilon = \varepsilon_{sm}
+ \varepsilon_\phi$, while the second generalized Friedmann equation (\ref{eq:SecondFriedEq}) takes on the form:
\eqn{ \frac{\ddot{a}}{a}
= \left[ -\frac{4\pi G_0}{c_0^2}\left(\varepsilon +3p\right) \right] e^{\phi/2\phi_0} + \frac{1}{4} \frac{\dot{\phi}}{\phi_0}\frac{\dot{a}}{a} \quad \quad (\Lambda=0),
}
where $p=p_{\phi}+p_{sm}$.
One way to derive the equation of motion for $\phi(t)$ is to set $\varepsilon_{sm}=p_{sm}=0$ and rewrite the continuity equation in
terms of $\phi(t)$. In fact, plugging
\eqn{ \dot{\varepsilon}_\phi = \frac{1}{2} \frac{1}{ c_0^2 e^{\phi/2\phi_0}}\left(-
\frac{1}{2} \frac{\dot{\phi}}{\phi_0}\dot{\phi}^2 + 2\dot{\phi}
\ddot{\phi}\right) + \dot{\phi} \frac{\partial V(\phi)}{\partial
\phi}, }
into the continuity equation (\ref{eq:ContinuityEq}),
gives the generalized Klein-Gordon equation\footnote{The very same equation of motion can be obtained by computing 
\eqn{ \frac{\delta S }{\delta \phi} = \frac{\delta
S_{gEH} }{\delta \phi}+\frac{\delta S_{vc} }{\delta \phi}=0, \nonumber
} which explicitly shows that the action principle is well-defined. We shall not consider couplings of the  $c$-field with standard matter because our goal is to investigate how $\phi$ could $c$-flate the universe; this is why the term $\frac{\delta S_{sm}}{\delta \phi}$ was not included in the expression for $\frac{\delta S}{\delta \phi}$.  }:
\eqn{\ddot{\phi} + 3 H \dot{\phi} - \frac{1}{4} \frac{\dot{\phi}^2}{\phi_0} + c_0^2e^{\phi/2\phi_0} \frac{\partial V(\phi)}{\partial \phi} = 0. \label{gfe_1}
}
This equation will be dealt with next.
%Note the presence of the third term on the
%LHS of the above equation of motion and the exponential coupling with
%the derivative of the potential. These terms modify the standard scalar
%field equation of motion and their physical consequences will be investigated
%in the next section.

%%%%% SUB-SECTION %%%%%

\subsection{How to $c$-flate the universe}

In the previous section, we showed the different kinematic ways $c$-flation framework solves the early universe puzzles and how it encompasses the different mechanisms found in the literature. We have now defined the dynamics of the scalar field that represents $c$ (or equivalently $G$). We can now discuss the physics that leads to those solutions of the SCM problems, and the condition on the field and its potential in order to do so.

 The equation of motion (\ref{gfe_1}) for the propagating degree of freedom $\phi$ is the same as:
\begin{equation}
 \ddot{\phi}+3H\dot{\phi}-\sqrt{\pi}\dot{\phi}^{2}+e^{2\sqrt{\pi}\phi}\frac{\partial V}{\partial\phi}=0\,, \label{eq:GKGPhi}
\end{equation}
where $\phi_{0}=\left(c_{0}^{4}/16\pi G_{0}\right)^{1/2}=1/4\sqrt{\pi}$, assuming $c_0=G_0=1$ and $\kappa_1=\kappa_2=1/2$.
The generalized Friedmann equations 
  (\ref{eq:FirstFriedEq}) and (\ref{eq:SecondFriedEq}) then read:
\begin{align}
& H^{2} =\frac{8\pi}{3}e^{2\sqrt{\pi}\phi}\varepsilon\,, \nonumber\\
& \frac{\ddot{a}}{a} = -4\pi e^{2\sqrt{\pi}\phi}\left(\varepsilon+3p\right)+\sqrt{\pi}\frac{\dot{a}}{a}\dot{\phi}\,, \label{eq:2ndFriedPhi}
\end{align}
where
\begin{equation}
\varepsilon = e^{-2\sqrt{\pi}\phi}\frac{\dot{\phi}^{2}}{2} + V \,, \qquad p = e^{-2\sqrt{\pi}\phi}\frac{\dot{\phi}^{2}}{2}-V\,,
\end{equation}
cf. Eq.~(\ref{energy_presure}). It is important to point out some differences with respect to the standard canonical scalar field in a cosmological background in GR. We observe that both Friedmann equations are different than usual, due to the
exponential enhancing factors. In addition, the second Friedmann equation (\ref{eq:2ndFriedPhi}) displays an unusual curvature-like term. Moreover, our generalized Klein-Gordon equation (\ref{eq:GKGPhi}) exhibits an extra friction term.
%\notere{(I don't agree with the statement that we have another friction term. We have an extra term proportional to $\dot{\phi}^2$ which is not a friction term. The original or the new version proposed by Rodrigo are fine for me once this point is sorted.} \gui{(that term is a damping term, so not sure why you don't agree)}

%\notere{*}Having determined the dynamical equations, we can now describe the physical conditions necessary to $c$-flation. We begin in Section \ref{sec:slow-roll} by considering the general case of slow-roll $c$-flation and show that it reduces to some specific known cases. \rod{\sout{We also show that this framework can provide a} In Section \ref{sec:PhaseTrans} we tend to the} solution \rod{\sout{for} $c$-flation provides to} the cosmological puzzles via a phase transition in $c$, similarly to the approach by VSL cases.\notere{(I suggest we remove this paragraph)*} \gui{(agree)}

%%%%%%%%%%%%%

\subsubsection{Slow-roll $c$-flation \label{sec:slow-roll}} 

We were able to describe $c$-flation in a similar way as the standard inflationary model, after checking that $d[(aH)^{-1}c]/dt < 0 \, \Rightarrow \omega < -1/3\, \Rightarrow \epsilon_c < 1\, \Rightarrow \ddot{a} > \dot{a} \dot{c}/c$. This is consistent with condition (\ref{eq:inflatinghorizon}),
\begin{equation} \label{eq:FirstSlowRollCond}
 \epsilon_c = - \frac{1}{c} \frac{(H/c)^{\cdot}}{(H/c)^2} = 4\pi \frac{\dot{\phi}^2}{H^2} \ll 1\,.
\end{equation}
Using the first Friedmann equation, this implies that $\dot{\phi^2}/e^{2\sqrt{\pi}\phi} < V(\phi)$, which means that the potential term dominates over the kinetic term, yielding
\begin{equation}
\left(\frac{H}{e^{\sqrt{\pi}\phi}}\right)^{2} \sim \frac{8\pi }{3 } V\,.
\label{H/c_const}
\end{equation}
Condition $\dot{\phi^2}/e^{2\sqrt{\pi}\phi} \ll V(\phi)$ should be satisfied. It is equivalent to the constraint $|\eta_c| \ll 1$ on the generalized second slow-roll parameter (\ref{eq:etac}),
\begin{equation}
\ddot{\phi} - \sqrt{\pi}\dot{\phi}^2  \ll \frac{e^{2\sqrt{\pi}\phi}}{2} V_{,\phi}\,, \label{eq:2ndSlowRollPhi}
\end{equation}
where we can define the parameter $\delta =2| \ddot{\phi} - \sqrt{\pi}\dot{\phi}^2| / e^{2\sqrt{\pi}\phi} V_{,\phi}$.  Eqs.~(\ref{eq:2ndSlowRollPhi}) and (\ref{eq:FirstSlowRollCond}) lead to the simplified equation of motion
\begin{equation}
3H\dot{\phi} \sim - e^{2\sqrt{\pi}\phi} V_{,\phi}\,.
\end{equation}
This is the equation of motion for the field $\phi$ during slow-roll $c$-flation. Given a potential, it determines the dynamics of $\phi$  that solves SCM problems.

Incidentally, the generalized slow-roll parameters $\epsilon_c$ and $\eta_c$ can be recast in terms of the potential $V$:
\begin{align}
& \epsilon_c^{V}= \frac{1}{16\pi} \left( \frac{V_{, \phi}}{V} \right)^2\,, \\
& |\eta_c^V| =  \frac{1}{4\pi} \frac{|V_{, \phi\, \phi}|}{V}\,,
\end{align}
where $\eta_c^V = \delta + \epsilon_c$. The conditions for shrinking the particle horizon and solving SCM problems are $\epsilon_c^V \ll 1$, and  $|\eta_c^V| \ll 1$.  As for inflation, a very flat potential can meet these conditions. 
Eq.~(\ref{H/c_const}) translates this requirement to $c$-flation framework: $(H/c)$ should be approximately constant. This is a generalization with respect to standard inflation, which demands $H \sim \mathrm{const.}$ On the other hand, slow-roll $c$-flation depends on the choice of both $c$ evolution and the shape of the potential $V$.

We can now comment on some possible scenarios. As pointed out in Section \ref{sec:SolvingPuzzles}, the conditions for a shrinking particle horizon does not necessarily imply that the associated period is accelerated. In fact, a proper choice of $V$ may set an evolution such that $\dot{c} < 0$ while $\dot{a} < 0$ yielding $\ddot{a} < 0$.
In general, this type of change in $c$ requires a smaller number of $c$-flation \textit{e-folds} than in standard inflation. We can see that by rewriting equation (\ref{eq:Cond2}): 
\begin{equation}
\Delta N - \sqrt{\pi} \Delta \phi > 64\,,
\label{Cond_2_rewritten}
\end{equation}
where we have considered $H = \mathrm{const.}$ for simplicity, and $\Delta N = \ln a_E - \ln a_I$.
If the initial value of the speed of light is larger than the value after $c$-flation, $c_I > c_E$, then $\Delta N > 64 - \sqrt{\pi} |\Delta \phi|$.
The opposite might also happen: if the speed of light was slower in the past, $c_E > c_I$, it would have caused the particle horizon to $c$-inflate at a slower rate thus requiring more time to resolve the early universe problems. 

We have explored $c$-flation general features and how it is able to solve the SCM problems. The specific mechanism for that depends on the choice of the potential for the related field. This is similar to what happens in the case of inflation. We will explore specific $c$-flation models as realizations of different potentials in future work.

%%%%%%%%%%%%%

\subsubsection{Phase Transition \label{sec:PhaseTrans}}

An interesting mechanism observed in VSL models  solving the early universe puzzles is a phase transition in $c$. This case can be realized in $c$-flation framework. We briefly show here how this can be implemented.

Changes in $c$ normalized by its value are able to drive $c$-flation. However, an instantaneous change in $c$ might also do the work. We can see that from
\begin{equation}
H \Delta t - \sqrt{\pi} \Delta \phi > 64\,,
\end{equation}
where, once again, we have taken $H = \mathrm{const.}$ for simplicity. A sharp change in $\phi$, with $\Delta t \sim 0$, leads to
\begin{equation}
\Delta \phi \sim - \frac{60}{\sqrt{\pi}}\,,
\end{equation}
which implies that $c_I \sim e^{60} c_E$. As a conclusion: the speed of light changes considerably in a small amount of time.

This is a very extreme case, similar to what is presented by VSL models. However, since now we have dynamics for $\phi$, the field phase transition can be realized given an appropriate choice of a symmetry breaking potential.

%%%%%%% SECTION %%%%%%%

\section{Discussion and Conclusions}\label{sec:conclusions}

We have used a covariant generalization of GR that allowed the variation of the
fundamental constants $c$, $G$ and $\Lambda$.
The cosmological background of this theory provides a set of generalized Friedmann
equations. Due to the general constraint, the set of three constants becomes two propagating degrees of freedom, or only one if $\Lambda=0$. We have formulated
the horizon and flatness problem in this framework and showed the necessary conditions to solve them after shrinking the comoving particle horizon. This condition implies an equation of state $w<-1/3$, the same requirement as in standard GR. However, we saw that it did not demand an obligatory accelerated expansion period, as in standard inflation. On the other hand, $c$-flation accommodates a decelerated expansion and even contraction phases obeying the conditions for horizon shrinking. This fact shows that $c$-flation framework encompasses many of the different mechanisms used in the literature for solving the initial condition problems of the early universe evolution.

Initially, we performed a kinematic analysis which required that two (of the three) ``constants'' depend on time in an \textit{ad hoc} manner.

However, the fact that $c$-flation framework preserves general covariance allowed us to write an action, where the varying constants enter as scalar fields and the general constraint is included via Lagrange multiplier. 
We have thus recovered Friedmann equations for the case $\Lambda=0$; these equations are identical to the ones derived within the kinematic approach. However, in the dynamical approach $c$ (or equivalently $G$) enters as a scalar field $\phi$. The equation of motion for $\phi$ was obtained.
We also derived the conditions a potential has to satisfy for producing a slow-roll $c$-flation phase and showed that this can lead to a superluminal evolution of $c$, solving SCM puzzles in a shorter number of \textit{e-folds} than traditional inflation. This might be important in order to avoid trans-Planckian effects that are present, for example, in some models of inflation \cite{Martin:2000xs,Brandenberger:2000wr} and that could affect the evolution of their perturbations
(see however \cite{Starobinsky:2001kn,Starobinsky:2002rp} for a proof that, as long as Lorentz symmetry is not broken by hand, trans-Planckian effects do not change the predictions of the inflationary power spectrum). 
We can also have a subluminal expansion leading to a longer period of $c$-flation. The dynamical approach via an action integral also gives a better-motivated way of realizing $c$-flation with a phase transition in $c$ by introducing a symmetry breaking potential. The details of these models for chosen potentials are not in the scope of this paper and are left for future
explorations.

The study of perturbations in this model is an important next step in the construction of covariant $c$-flation models, as one wants to compare its predictions with CMB observations and large scale structure (LSS) surveys of the universe. Our framework should be able to reproduce the observed nearly scale-invariant power spectrum through the
dynamics of the field which will strongly depend on the choice of the potential. 
It would be particularly interesting to determine any observations that might differentiate
our framework from inflation and bouncing models. For example, if $c$-flation occurred as a superluminal expansion and lasted less \textit{e-folds} than the usual inflation, this should yield distinct effects on the CMB and the LSS observables. As noticed in
\cite{Afshordi:2016guo} in the context of bimetric models, an alternative mechanism in the early universe might give a different form for the expected
spectral index of curvature perturbations and a variant to the gravitational waves spectrum. This may also be the case if a phase transition occurs.

One might notice that the action given by our theory is general and potentially describes several cosmological effects. An interesting and straightforward example is to model dark energy using a mechanism
analogous to $c$-flation thus generating the current acceleration of
the universe. This is a work in progress.

%%%%%%%%%%%%%%%%%%%%%%%%%%%

\begin{acknowledgments}
We thank Robert Brandenberger for useful discussions and the anonymous referee for the suggestions to improve the text. E.\,F. and
G.\,F. acknowledge financial support from CNPq (Science Without Borders).
R.C. thanks financial support by the SARChI NRF grant holder. R.C. thanks McGill University
for additional hospitality and partial financial support during the beginning of this project. R.R.C.
is grateful to CAPES-Brazil for partial financial support.
\end{acknowledgments}

%%%%%%%%%%%%%%%%%%%%%%%%%%%%

\appendix

\section{Including $\Lambda(t)$}\label{sec:A}

We have briefly mentioned that $\Lambda (t)$ could solve the early universe problems, but we have not shown how that would be the case within our covariant framework. Let us explore this possibility now. 

%%%%% SUB-SECTION %%%%%

\subsection{Flatness Problem}

In this case, the first Friedmann equation bears a term containing
$\Lambda(t)$, as in (\ref{eq:FirstFriedEq}). It can also be
written as,
\begin{equation}
\varepsilon-\left(\frac{3c^{2}}{8\pi G}H^{2}-\frac{3c^{4}}{8\pi G}\frac{\Lambda}{3}\right)=\frac{3c^{4}}{8\pi G}\frac{k}{a^{2}}\,. \label{eq:epsilon(Lambda)}
\end{equation}
The critical energy density $\varepsilon_c$ is defined to be the one rendering $k=0$. From Eq.~(\ref{eq:epsilon(Lambda)}), this corresponds to set:
\begin{equation}
\varepsilon_{c}\equiv\left(\frac{3c^{2}}{8\pi G}H^{2}-\frac{3c^{4}}{8\pi G}\frac{\Lambda}{3}\right)=\left(\frac{3c^{2}}{8\pi G}H^{2}-\varepsilon_{\Lambda}\right)\,;\qquad\varepsilon_{\Lambda}\equiv\frac{\Lambda c^{4}}{8\pi G}\,.\label{eq:epsilon_c(Lambda)}
\end{equation}
We use the definitions
\begin{equation}
\epsilon_{\Omega}=\Omega-1\:;\qquad\Omega=\frac{\varepsilon}{\varepsilon_{c}}\, ,\label{eq:Omega}
\end{equation}
from which it follows the identity,
\begin{equation}
\dot{\epsilon}_{\Omega}=\left(1+\epsilon_{\Omega} \right)\left(\frac{\dot{\varepsilon}}{\varepsilon}-\frac{\dot{\varepsilon}_{c}}{\varepsilon_{c}}\right)\,.\label{eq:epsilon_dot}
\end{equation}
After tedious manipulations involving the definition (\ref{eq:epsilon_c(Lambda)}) and the second Friemdann equation (\ref{eq:SecondFriedEq}), the equation for $\left(\dot{\varepsilon}_{c}/\varepsilon_{c}\right)$
is,
\[
\frac{\dot{\varepsilon}_{c}}{\varepsilon_{c}}=-\left[\left(\frac{\dot{G}}{G}-4\frac{\dot{c}}{c}\right)+3\frac{\dot{a}}{a}\left(1+\frac{p}{\varepsilon}\right)+\epsilon_{\Omega} \frac{\dot{a}}{a}\left(1+3\frac{p}{\varepsilon}\right)\right]-\frac{\varepsilon_{\Lambda}}{\varepsilon_{c}}\left(\frac{\dot{\Lambda}}{\Lambda}\right)\,.
\]
Interestingly enough, the first term and the last are related through
the GC (\ref{eq:ConstraintEq}) in the form,
\begin{equation}
-\left(\frac{\dot{G}}{G}-4\frac{\dot{c}}{c}\right)=\frac{\varepsilon_{\Lambda}}{\varepsilon}\left(\frac{\dot{\Lambda}}{\Lambda}\right)\,.\label{eq:ConstraintEq(epsilonL)}
\end{equation}
Hence, we finally get,
\begin{equation}
\dot{\epsilon}_{\Omega} = \left(1+\epsilon_{\Omega} \right) \epsilon_{\Omega} \frac{\dot{a}}{a} \left(1+3\frac{p}{\varepsilon}\right) + \left(1+\epsilon_{\Omega} \right)\epsilon_{\Omega} \,\frac{\varepsilon_{\Lambda}}{\varepsilon}\left(\frac{\dot{\Lambda}}{\Lambda}\right)\,.\label{eq:epsilon_L}
\end{equation}
This equation should be compared to the result due to Albrecht and
Magueijo (A\&M) \cite{AM}, namely
\begin{equation}
\dot{\epsilon}_{\Omega} = \left(1+\epsilon_{\Omega} \right) \epsilon_{\Omega} \frac{\dot{a}}{a}\left(1+3\frac{p}{\varepsilon}\right) + 2\frac{\dot{c}}{c}\epsilon_{\Omega}\,. \qquad(\text{A\&M})\label{eq:epsilon_c_dot(AM)}
\end{equation}
The difference is in the last term of Eqs.~(\ref{eq:epsilon_L}-\ref{eq:epsilon_c_dot(AM)}).

In the SCM scenario $\dot{\Lambda}=0$ and Eq. (\ref{eq:epsilon_L}) reduces to:
\[
\dot{\epsilon}_{\Omega} =\left(1+\epsilon_{\Omega} \right) \epsilon_{\Omega} \frac{\dot{a}}{a}\left(1+3\frac{p}{\varepsilon}\right)\,.
\]
Thus, in an expanding universe $\left(\frac{\dot{a}}{a}>0\right)$
mantaining the strong energy condition $\left(1+3\frac{p}{\varepsilon}\right)>0$
we shall always have $\dot{\epsilon}_{\Omega} > 0$. This means that $\epsilon_{\Omega}$
grows with cosmic time and we cannot explain the observational fact
that $\epsilon_{\Omega} \simeq 0$ today unless we impose a huge fine tuning
as initial condition.

In fact, radiation is consistent with $p=\varepsilon/3$, \textit{i.e.}, in SCM this implies $\epsilon_{\Omega} \sim a^{2}$; for
matter, $p\simeq0$ so that $\epsilon_{\Omega} \sim a$. This leads to a total growth
of 32 orders of magnitude since the Planck epoch \cite{AM}. Therefore, one requires
$\epsilon_{\Omega} < 10^{-32}$ at $t=t_{P}$ if one considers a radiation dominated phase. This is just another phrasing of the flatness problem. In \cite{AM} it is  solved by admitting an early sharp
phase transition in which $\left|\dot{c}/c\right|\gg\dot{a}/a$ and
$\left(\dot{c}/c\right)<0$; then Eq. (\ref{eq:epsilon_c_dot(AM)}) lead to
\begin{equation}
\epsilon_{\Omega} \sim c^{2}\,,\qquad(\text{A\&M}) \label{eq:epsilon(c)AM}
\end{equation}
and $\epsilon_{\Omega} \simeq 0$ is achieved today for a $\epsilon_{\Omega} \approx 1$ around
or before the phase transition.

 In our case, we consider
\begin{equation}
\frac{\varepsilon_{\Lambda}}{\varepsilon}\left(\frac{\dot{\Lambda}}{\Lambda}\right)\gg\frac{\dot{a}}{a}\,, \label{eq:Gdot_gg_adot}
\end{equation}
then (\ref{eq:epsilon_L}) reads
\begin{equation}
\dot{\epsilon}_{\Omega} \approx \left(1+\epsilon_{\Omega} \right) \epsilon_{\Omega} \,\frac{\varepsilon_{\Lambda}}{\varepsilon}\left(\frac{\dot{\Lambda}}{\Lambda}\right)\,,\label{eq:epsilon_c_dot(Lambda)}
\end{equation}
and $\epsilon_{\Omega} \simeq 0$ may be achieved nowadays if $\Lambda$ was a decaying function of
time for long enough, without the need of any exotic equation of state in the expanding universe.

%%%%% SUB-SECTION %%%%%

\subsection{Cosmological Constant Problem}

What we regard as the cosmological constant problem does not concern
its very existence but rather the fine tuning required to explain
its present-day observational value. This will become clearer below.

By differentiating $\varepsilon_{\Lambda}$ \textendash{} defined in Eq.~(\ref{eq:epsilon_c(Lambda)}) \textendash{} one gets:
\[
\frac{\dot{\varepsilon}_{\Lambda}}{\varepsilon_{\Lambda}}=\left(\frac{\dot{\Lambda}}{\Lambda}\right)-\left(\frac{\dot{G}}{G}-4\frac{\dot{c}}{c}\right)\,.
\]
In view of (\ref{eq:ConstraintEq(epsilonL)}), this result is
equivalent to,
\begin{equation}
\frac{\dot{\varepsilon}_{\Lambda}}{\varepsilon_{\Lambda}}=\left(\frac{\dot{\Lambda}}{\Lambda}\right)\left(1+\frac{\varepsilon_{\Lambda}}{\varepsilon}\right)\,.\label{eq:varepsilonL_dot}
\end{equation}

Following Albrecht \& Magueijo\footnote{Actually, they write $\epsilon_{\Lambda}=\rho_{\Lambda}/\rho_{m}$
above their equation (15).}, we define,
\begin{equation}
\epsilon_{\Lambda}=\frac{\varepsilon_{\Lambda}}{\varepsilon}\,.\label{eq:epsilonL}
\end{equation}
Then,
\[
\dot{\epsilon}_{\Lambda}=\epsilon_{\Lambda}\left(\frac{\dot{\varepsilon}_{\Lambda}}{\varepsilon_{\Lambda}}-\frac{\dot{\varepsilon}}{\varepsilon}\right)\,.
\]
The term $\left(\dot{\varepsilon}_{\Lambda}/\varepsilon_{\Lambda}\right)$
in the right hand side was calculated in (\ref{eq:varepsilonL_dot}). The
term $\left(\dot{\varepsilon}/\varepsilon\right)$ is given by
 (\ref{eq:ContinuityEq}). Therefore, the
equation for $\dot{\epsilon}_{\Lambda}$ is put in the form:
\begin{equation}
\dot{\epsilon}_{\Lambda}=\epsilon_{\Lambda}\left[3\frac{\dot{a}}{a}\left(1+\frac{p}{\varepsilon}\right)+\left(\frac{\dot{\Lambda}}{\Lambda}\right)\left(1+\epsilon_{\Lambda}\right)\right]\,.\label{eq:epsilonL_dot}
\end{equation}
This result is analogous to Eq.~(15) in the paper \cite{AM} by Albrecht \& Magueijo, viz.
\begin{equation}
\dot{\epsilon}_{\Lambda}=\epsilon_{\Lambda}\left[3\frac{\dot{a}}{a}\left(1+\frac{p}{\varepsilon}\right)+2\left(\frac{\dot{c}}{c}\right)\frac{\left(1+\epsilon_{\Lambda}\right)}{\left(1+\epsilon_{\Omega} \right)}\right]\qquad(\text{A\&M}),\label{eq:epsilonL_dot(AM)}
\end{equation}
but the last term in both equations are, of course, different.

Now we can clarify the meaning of the fine tuning problem of cosmological
constant. In SCM, $\dot{\Lambda}=0$ and the equation
above for $\dot{\epsilon_{\Lambda}}\left(\dot{\Lambda}\right)$ can
be integrated for both radiation and matter giving $\epsilon_{\Lambda}\sim a^{4}$
and $\epsilon_{\Lambda}\sim a^{3}$ respectively. This means a total
grow of 64 orders of magnitude since Planck time assuming radiation-domination; this very fact makes
it difficult to explain why $\epsilon_{\Lambda}\approx1$ today.

Some comments are in order:
\begin{enumerate}
\item A\&M are able to solve de $\Lambda$-problem through the term scaling
with $\left(\dot{c}/c\right)$ in the equation for $\dot{\epsilon}_{\Lambda}$;
we are not able to do so directly because $\Lambda$ and its time-derivative
$\dot{\Lambda}$ appear in the very expression for $\dot{\epsilon}_{\Lambda}$.
However, our framework makes room for the simultaneous interplay between
$\Lambda$ and varying $c$ and $G$.

\item From (\ref{eq:epsilonL_dot}) we see that we can solve the cosmological problem if we have a phase in which
\begin{equation}
\left|\left(\frac{\dot{\Lambda}}{\Lambda}\right)\right| \ll \frac{\dot{a}}{a},
\end{equation}
where $\Lambda$ decreases with time. As long as this phase lasts long enough, we would have alleviated the cosmological problem defined above, as well as solved the flatness problem. A similar argument applies to the horizon problem.
\end{enumerate}

%%%%%%% SECTION %%%%%%%

\section{A second generalized Einstein-Hilbert action} \label{sec:B}

%%%% Rodrigo: revise from here!!!

In this appendix, we show explicitly how the vanilla single field inflation scenario can be interpreted as a subset of the more general set up of $c$-flation.

The basic hypothesis to derive inflation from the $c$-flation scenario is to consider that the causality speed $c_{ST}$ (the one that appears in the metric) and the Einstein speed $c_E$ (the one that appears on the right hand side of Einstein's equation) are different. More precisely, we require that $c_E = c_E(x^\alpha)$ is varying and $c_{ST}=c_0$ is a constant and coincide with the speed of light in vacuum.

By setting $\Lambda=0$, solving the GC and plugging it back into action (\ref{eq:action}) gives:
\eqn{\label{3.6}
S_{vc}= \frac{c_0^4}{16\pi G_0}\int d^4 x \sqrt{-g} \left( -\frac{(\kappa_1+\kappa_2)}{2
\Phi^2} \nabla_\mu \Phi \nabla^\mu \Phi - V(\Phi) \right).
}
Note that the metric does not depend on $\Phi$. After a field redefinition identical to (\ref{eq:fieldredefinition}), this action becomes\rod{:}
\eqn{\label{3.6}
S_{vc}= \int d^4 x \sqrt{-g} \left( -\frac{1}{2}\nabla_\mu \phi \nabla^\mu \phi - V(\phi) \right).
}
The action describing the full system in this case is:
\eqn{\label{inf}
S = \int d^4 x \sqrt{-g}\left( \frac{c_0^4}{16\pi G_0} R -\frac{1}{2}\nabla_\mu \phi \nabla^\mu \phi - V(\phi)\right) + S_{sm}.
}

With a constant $c_{ST}$, the gravitational sector is not modified, and we get the well known vanilla inflationary picture. In our framework, to consider a period where the two speeds were different from each other is not a mere conceptual subject, but rather has an impact on the evolution of the universe. It can solve SCM puzzles and provide a different understanding of what the inflaton actually is.

%%%%%%%%%%%%%%%%%%%%%%%%%%%%%%%%%%%%%%%%%%%%%%%%%%%%%%%%%%%%%%%%%%%%%%%%%%%%%%%%%%%


\begin{thebibliography}{10}


\bibitem{Starobinsky:1980te} A.~A.~Starobinsky, %``The Inflationary Universe: A Possible Solution to the Horizon and Flatness Problems,''
 Phys. Lett.\ B \textbf{91}, 99-102 (1980) %doi:10.1103/PhysRevD.23.347
 \href{https://www.sciencedirect.com/science/article/abs/pii/037026938090670X?via}{10.1016/0370-2693(80)90670-X}.

\bibitem{Guth:1980zm} A.~H.~Guth, %``The Inflationary Universe: A Possible Solution to the Horizon and Flatness Problems,''
 Phys.\ Rev.\ D \textbf{23}, 347 (1981) %doi:10.1103/PhysRevD.23.347
 \href{https://journals.aps.org/prd/pdf/10.1103/PhysRevD.23.347}{10.1103/PhysRevD.23.347}.

\bibitem{Linde:1981mu} A.~D.~Linde, %``A New Inflationary Universe Scenario: A Possible Solution of the Horizon, Flatness, Homogeneity, Isotropy and Primordial Monopole Problems,''
 Phys.\ Lett.\ \textbf{108B}, 389 (1982) \href{http://ac.els-cdn.com/0370269382912199/1-s2.0-0370269382912199-main.pdf?_tid=1f16a6d8-29d4-11e7-a4a4-00000aacb360&acdnat=1493137849_ae074adc8a976f612137dca8d18bc7d2}{10.1016/0370-2693(82)91219-9}.

\bibitem{Albrecht:1982wi} A.~Albrecht and P.~J.~Steinhardt, %``Cosmology for Grand Unified Theories with Radiatively Induced Symmetry Breaking,''
 Phys.\ Rev.\ Lett.\ \textbf{48}, 1220 (1982) %doi:10.1103/PhysRevLett.48.1220
\href{https://journals.aps.org/prl/pdf/10.1103/PhysRevLett.48.1220}{10.1103/PhysRevLett.48.1220}.


\bibitem{Starobinsky:1979ty}  A.~A.~Starobinsky,
%``Spectrum of relict gravitational radiation and the early state of the universe,''
 JETP Lett.\ \textbf{30}, 682-685 (1979) 
 %{[}Pisma Zh.\ Eksp.\ Teor.\ Fiz.\ \textbf{33},
549 (1981){]}
\href{http://www.jetpletters.ac.ru/ps/1370/article_20738.pdf}{JETP Lett.30}.


\bibitem{Mukhanov:1981xt} V.~F.~Mukhanov and G.~V.~Chibisov,
%``Quantum Fluctuations and a Nonsingular Universe,''
 JETP Lett.\ \textbf{33}, 532 (1981) 
 %{[}Pisma Zh.\ Eksp.\ Teor.\ Fiz.\ \textbf{33},
%549 (1981){]}
\href{http://www.jetpletters.ac.ru/ps/1510/article_23079.pdf}{JETP Lett.33}.

\bibitem{Ade:2013zuv} P.~A.~R.~Ade \textit{et al.} {[}Planck Collaboration{]},
%``Planck 2013 results. XVI. Cosmological parameters,''
 Astron.\ Astrophys.\ \textbf{571}, A16 (2014) %doi:10.1051/0004-6361/201321591
\href{https://arxiv.org/pdf/1303.5076.pdf}{[astro-ph/1303.5076]}.

\bibitem{Baumann:2009ds} D.~Baumann, \textit{TASI lectures on inflation, }%``Inflation,''
 %doi:10.1142/9789814327183$_$0010
\href{https://arxiv.org/pdf/0907.5424.pdf}{[hep-th/0907.5424]}.

\bibitem{Brandenberger:2012zb} R.~H.~Brandenberger, %``The Matter Bounce Alternative to Inflationary Cosmology,''
\href{https://arxiv.org/pdf/1206.4196.pdf}{[astro-ph/1206.4196]}.

\bibitem{Khoury:2001wf} J.~Khoury, B.~A.~Ovrut, P.~J.~Steinhardt
and N.~Turok, %``The Ekpyrotic universe: Colliding branes and the origin of the hot big bang,''
 Phys.\ Rev.\ D \textbf{64}, 123522 (2001) %doi:10.1103/PhysRevD.64.123522
\href{https://arxiv.org/pdf/hep-th/0103239.pdf}{[hep-th/0103239]}.

\bibitem{Buchbinder:2007ad} E.~I.~Buchbinder, J.~Khoury and B.~A.~Ovrut,
%``New Ekpyrotic cosmology,''
 Phys.\ Rev.\ D \textbf{76}, 123503 (2007) %doi:10.1103/PhysRevD.76.123503
\href{https://arxiv.org/pdf/hep-th/0702154.pdf}{[hep-th/0702154]}.

\bibitem{Gasperini:1992em} M.~Gasperini and G.~Veneziano, %``Pre - big bang in string cosmology,''
 Astropart.\ Phys.\ \textbf{1}, 317 (1993) %doi:10.1016/0927-6505(93)90017-8
\href{https://arxiv.org/pdf/hep-th/9211021.pdf}{[hep-th/9211021]}.

\bibitem{Thomson_Taid} W.~Thomson and P.~G.~Taid, \textit{Treatise
in natural phylosophy}, University Press (1879).

\bibitem{Dirac} P.~A.~M. Dirac, %"The Cosmological Constants,"
 Nature \textbf{139}, 323 (1937). %doi:10.1038/139323a0

\bibitem{Uzan:2002vq} J.~P.~Uzan, %``The Fundamental constants and their variation: Observational status and theoretical motivations,''
 Rev.\ Mod.\ Phys.\ \textbf{75}, 403 (2003) %doi:10.1103/RevModPhys.75.403
\href{https://arxiv.org/pdf/hep-ph/0205340.pdf}{[hep-ph/0205340]}.



\bibitem{Padmanabhan:2016lul} T.~Padmanabhan, %``Do We Really Understand the Cosmos?,''
\href{https://arxiv.org/pdf/1611.03505.pdf}{[gr-qc/1611.03505]}.

\bibitem{Riess} A.~G.~Riess \textit{et al.} {[}Supernova Search
Team{]}, %``Observational evidence from supernovae for an accelerating universe and a cosmological constant,''
 Astron.\ J.\ \textbf{116}, 1009 (1998)
 \href{https://arxiv.org/pdf/astro-ph/9805201.pdf}{[astro-ph/9805201]}.

\bibitem{Perlmutter} S.~Perlmutter \textit{et al.} {[}Supernova
Cosmology Project Collaboration{]}, %``Measurements of Omega and Lambda from 42 high redshift supernovae,''
 Astrophys.\ J.\ \textbf{517}, 565 (1999) %doi:10.1086/307221 {[}astro-ph/9812133{]}.
 \href{https://arxiv.org/pdf/astro-ph/9812133.pdf}{[astro-ph/9812133]}.


\bibitem{Moffat93} J. W. Moffat, 
%``Superluminary universe: A possible solution to the initial value problem in cosmology,'' 
Int. J. Mod. Phys. D 2, 351 (1993) 
% doi:10.1142/S0218271893000246 
\href{https://arxiv.org/pdf/gr-qc/9211020.pdf}{[gr-qc/9211020]}.

\bibitem{AM} A.~Albrecht and J.~Magueijo, %``A Time varying speed of light as a solution to cosmological puzzles,''
 Phys.\ Rev.\ D \textbf{59}, 043516 (1999) %doi:10.1103/PhysRevD.59.043516
\href{https://arxiv.org/pdf/astro-ph/9811018.pdf}{[astro-ph/9811018]}.

\bibitem{Clayton:1998hv} M.~A.~Clayton and J.~W.~Moffat, %``Dynamical mechanism for varying light velocity as a solution to cosmological problems,''
 Phys.\ Lett.\ B \textbf{460}, 263 (1999) %doi:10.1016/S0370-2693(99)00774-1
\href{https://arxiv.org/pdf/astro-ph/9812481.pdf}{[astro-ph/9812481]}.

\bibitem{Bassett:2000wj} B.~A.~Bassett, S.~Liberati, C.~Molina-Paris
and M.~Visser, %``Geometrodynamics of variable speed of light cosmologies,''
 Phys.\ Rev.\ D \textbf{62}, 103518 (2000) %doi:10.1103/PhysRevD.62.103518
\href{https://arxiv.org/pdf/astro-ph/0001441.pdf}{[astro-ph/0001441]}.

\bibitem{BM} J.~D.~Barrow and J.~Magueijo, %``Varying alpha theories and solutions to the cosmological problems,''
 Phys.\ Lett.\ B \textbf{443}, 104 (1998) %doi:10.1016/S0370-2693(98)01294-5
\href{https://arxiv.org/pdf/astro-ph/9811072.pdf}{[astro-ph/9811072]}.

\bibitem{B} J.~D.~Barrow, %``Cosmologies with varying light speed,''
\href{https://arxiv.org/pdf/astro-ph/9811022.pdf}{[astro-ph/9811022]}.

\bibitem{higher_dim} E.~Kiritsis, %``Supergravity, D-brane probes and thermal superYang-Mills: A Comparison,''
 JHEP \textbf{9910}, 010 (1999) %doi:10.1088/1126-6708/1999/10/010
\href{https://arxiv.org/pdf/hep-th/9906206.pdf}{ [hep-th/9906206]}. S.~H.~S.~Alexander, %``On the varying speed of light in a brane induced FRW universe,''
 JHEP \textbf{0011}, 017 (2000) %doi:10.1088/1126-6708/2000/11/017
\href{https://arxiv.org/pdf/hep-th/9912037.pdf}{ [hep-th/9912037]}.

\bibitem{Avelino:1999is} P.~P.~Avelino and C.~J.~A.~P.~Martins,
%``Does a varying speed of light solve the cosmological problems?,''
 Phys.\ Lett.\ B \textbf{459}, 468 (1999) %doi:10.1016/S0370-2693(99)00694-2
\href{https://arxiv.org/pdf/astro-ph/9906117.pdf}{ [astro-ph/9906117]}.

\bibitem{BM2} J.~D.~Barrow and J.~Magueijo, %``Solutions to the quasi-flatness and quasilambda problems,''
 Phys.\ Lett.\ B \textbf{447}, 246 (1999) %doi:10.1016/S0370-2693(99)00008-8
\href{https://arxiv.org/pdf/astro-ph/9811073.pdf}{ [astro-ph/9811073]}.

\bibitem{EU} G.~F.~R.~Ellis and J.~P.~Uzan, %```c' is the speed of light, isn't it?,''
 Am.\ J.\ Phys.\ \textbf{73}, 240 (2005) 
%doi:10.1119/1.1819929
\href{https://arxiv.org/pdf/gr-qc/0305099.pdf}{ [gr-qc/0305099]}.

\bibitem{BelCar} J.~A.~Belinchon and J.~L.~Carames, %``A New formulation of a naive theory of time-varying constants,''
 Spacetime and Substance \textbf{6}, 97 (2005) 
\href{https://arxiv.org/pdf/gr-qc/0407068.pdf}{ [gr-qc/0407068]}.

\bibitem{Gui} G. Franzmann,  \href{https://arxiv.org/pdf/1704.07368.pdf}{ [gr-qc/1704.07368]}.

\bibitem{Carroll} S. Carroll, \textit{Spacetime and Geometry. An
Introduction to general Relativity}, Addison Wesley, San Francisco,
2004.

\bibitem{Ellis:2007ah} 
  G.~F.~R.~Ellis,
  %``Note on Varying Speed of Light Cosmologies,''
  Gen.\ Rel.\ Grav.\  {\bf 39}, 511 (2007)
  %doi:10.1007/s10714-007-0396-4
  \href{https://arxiv.org/pdf/astro-ph/0703751.pdf}{ [astro-ph/0703751]}.


\bibitem{Moffat06} J. W. Moffat, 
%``Scalar-Tensor-Vector Gravity Theory'' 
JCAP \textbf{03}, 004 (2006) 
% doi:10.1088/1475-7516/2006/03/004 
\href{https://arxiv.org/pdf/gr-qc/0506021.pdf}{[gr-qc/0506021]}.

\bibitem{Moffat16} J. W. Moffat, 
%``Variable speed of light cosmology, primordial fluctuations and gravitational waves,'' 
Eur. Phys. J. C \textbf{76} 130 (2016)  
% doi:10.1140/epjc/s10052-016-3971-6 
\href{https://arxiv.org/pdf/1404.5567.pdf}{[astro-ph/1404.5567]}.

\bibitem{Ayuso:2014jda} I.~Ayuso, J.~Beltran Jimenez, A.~de la
                        Cruz-Dombriz, %``TConsistency of universally nonminimally coupled
                    %    $f(R,T,R_{\mu \nu}T^{\mu \nu})$ theories,''
 Phys.\ Rev.\ D \textbf{91}, no. 10, 104003 (2015) %doi:10.1103/PhysRevD.94.101301.
\href{https://arxiv.org/pdf/1603.03312.pdf}{{[}arXiv:1411.1636 {[}hep-th{]}{]}}.

\bibitem{Sami:2002se}M.~Sami, T.~Padmanabhan, %``A Viable cosmology with a scalar field coupled to the
                        %trace of the stress tensor,''
 Phys.\ Rev.\ D \textbf{67}, 083509 (2003) %doi:10.1103/PhysRevD.94.101301.
\href{https://arxiv.org/pdf/1603.03312.pdf}{{[}arXiv:hep-th/0212317{]}}.

\bibitem{Joyce:2014kja}A.~Joyce, B.~Jain, J.~Khoury, M.~Trodden, %`Beyond the Cosmological Standard Model,''
 Phys. Rept. \textbf{568}, 1-98 (2015) %doi:10.1103/PhysRevD.94.101301.
\href{https://arxiv.org/pdf/1603.03312.pdf}{{[}arXiv:1407.0059{[}astro-ph{]}{]}}.




\bibitem{Martin:2000xs} 
  J.~Martin and R.~H.~Brandenberger,
  %``The TransPlanckian problem of inflationary cosmology,''
  Phys.\ Rev.\ D {\bf 63}, 123501 (2001)
  %doi:10.1103/PhysRevD.63.123501
  \href{https://arxiv.org/pdf/hep-th/0005209.pdf}{ [hep-th/0005209]}.
  
 \bibitem{Brandenberger:2000wr} 
  R.~H.~Brandenberger and J.~Martin,
  %``The Robustness of inflation to changes in superPlanck scale physics,''
  Mod.\ Phys.\ Lett.\ A {\bf 16}, 999 (2001)
  %doi:10.1142/S0217732301004170
  \href{https://arxiv.org/pdf/astro-ph/0005432.pdf}{ [astro-ph/0005432]}.
  


  \bibitem{Starobinsky:2001kn} 
  A.~A.~Starobinsky, 
  %``Robustness of the inflationary perturbation spectrum to transPlanckian physics,''
  JETP Lett.\ \textbf{73}, 415-418 (2001) %{[}Pisma Zh.Eksp.Teor.Fiz. 73 (2001) 415-418, JETP Lett. 73 (2001) 371-374 {]}
  %doi:10.1134/1.1381588
  \href{https://arxiv.org/abs/astro-ph/0104043}{ [astro-ph/0104043]}.
  
 
   \bibitem{Starobinsky:2002rp} 
  A.~A.~Starobinsky and I.~I.~ Tkachev,
  %``Trans-Planckian particle creation in cosmology and ultra-high energy cosmic rays,''
  JETP Lett.\ \textbf{76}, 235-239 (2002) %{[}Pisma Zh.Eksp.Teor.Fiz. 76 (2002) 291-295 {]}
  %doi:10.1134/1.1520612
  \href{https://arxiv.org/abs/astro-ph/0207572}{ [astro-ph/0207572]}.

\bibitem{Afshordi:2016guo} N.~Afshordi and J.~Magueijo, %``The critical geometry of a thermal big bang,''
 Phys.\ Rev.\ D \textbf{94}, no. 10, 101301 (2016) %doi:10.1103/PhysRevD.94.101301.
\href{https://arxiv.org/pdf/1603.03312.pdf}{{[}arXiv:1603.03312 {[}gr-qc{]}{]}}.




\end{thebibliography}
\end{document}